\UseRawInputEncoding
\documentclass[pra,preprint,amssymb,showpacs]{revtex4}
\usepackage{dcolumn}
\usepackage{bm}
\usepackage{graphicx}
\def\be{\begin{equation}}
\def\ee{\end{equation}}
\def\bea{\begin{eqnarray}}
\def\eea{\end{eqnarray}}
\def\source#1#2#3#4{{\it #1}~{\bf #2}, #3 (#4)}
\def\Eq#1{Eq. \ref{#1}}
\def\Eqs#1#2{Eqs. \ref{#1} and \ref{#2}}
\def\Ref#1{Ref. \cite{#1}}

\def\ie{{\it i.e.}}
\def\eg{{\it eg.}}

\def\Sch{Schr{\"o}dinger}

\def\ket#1{| #1 \rangle}

\def\bra#1{\langle #1 |}
\def\braket#1#2{\langle #1| #2 \rangle}

\def\muup{\mu_\uparrow}
\def\mudown{\mu_\downarrow}

\def\Eup{E_\uparrow}
\def\Edown{E_\downarrow}
\def\Dk{D_k}
\def\Dup{D_{\uparrow}}
\def\Ddown{D_{\downarrow}}

\def\ZAup{Z_{A\uparrow}}
\def\ZAdown{Z_{A\downarrow}}
\def\XAup{X_{A\uparrow}}
\def\XAdown{X_{A\downarrow}}
\def\ZPk{Z_{Pk}}
\def\ZPup{Z_{P\uparrow}}
\def\ZPdown{Z_{P\downarrow}}

\def\CP{{\cal P}}

\begin{document}
\title{Observing a Quantum Measurement}
\author{Jay Lawrence} 
\affiliation{Department of Physics and Astronomy, Dartmouth
          College, Hanover, NH 03755, USA}
\affiliation{The James Franck Institute, University of Chicago, 
          Chicago, IL 60637}
\date{revised \today}
\bigskip
\begin{abstract}

With the example of a Stern-Gerlach measurement
on a spin-1/2 atom, we show that a superposition of both paths may be observed 
compatibly with properties attributed to state collapse - for example, the {\it singleness}
(or mutual exclusivity) of outcomes.  This is done by inserting a quantum two-state 
system (an ancilla) in each path, capable of responding to the passage of the atom, and 
thus acting as a virtual detector.  We then consider real measurements on the compound 
system of atomic spin and two ancillae.  Nondestructive measurements of a set of 
compatible joint observables can be performed, one for a superposition and others 
for collapse properties.  A novel perspective is given as to why, within unitary 
quantum theory, ordinary measurements are blind to such superpositions.  
Implications for the theory of measurement are discussed.
\end{abstract}
\maketitle
\section{Introduction}

Opinions differ on whether or not there is a quantum measurement problem, and if so, 
exactly what it is \cite{Schlossbook,survey}.  A more focused question regards the 
collapse of the state vector in projective measurements, as formalized by von 
Neumann \cite{VN55}.  By collapse we mean the observed process described by
one of the textbook postulates, which states (in its simplest form) \cite{preMGM,MGM}:

\noindent{{\it Measurement Postulate:  A measurement of the observable A yields 
one of its eigenvalues, $a_n$, and, in an ideal measurement  \cite{Ideal,Pokorny}, 
places the measured object in the corresponding eigenstate, $\phi_n$ (where 
$A \phi_n = a_n\phi_n$)}}. 

\noindent Included with this, or stated as a separate postulate, is the Born rule
probability of this outcome, $|\langle \phi_n | \psi \rangle |^2$, where $\psi$ is the
initial state of the object being measured.  All interpretations must agree on the above
as a statement of fact, but they disagree on its status - \ie, is it independent of the 
other postulates, which stipulate unitary evolution in the appropriate Hilbert space,
or is it derivable from them?   

Regarding attitudes on this more focused issue, it seems reasonable to identify a 
small number of broad categories.  Here is a grouping into three:  The most 
conservative takes collapse as axiomatic - that is, it cannot be derived from the 
other axioms - suggesting that the collapse process itself is not subject to quantum 
analysis \cite{Bohr1958,Weinberg1}.  This position is  consistent with most textbooks 
written over more than the last half century, which list collapse among the axioms.  It 
is intended to include those who apply quantum theory according to these textbook 
axioms, without (however) adopting any particular interpretation philosophically.  It 
also includes the epistemic and information-based approaches \cite{Cabellomap},
whose intellectual roots extend back to the Copenhagen interpretation 
\cite{Bohr1963,Wigner1961}.  We refer to this general position as Standard 
Quantum Theory (SQT), interpreted broadly.

A contrasting position is that the collapse phenomenon, as observed, is in fact derived 
from the other axioms, following unitary evolution of an appropriate closed system 
which includes the apparatus and the relevant environment \cite{relevant,Zurek.81}, 
as well as the object of study.  And indeed unitary evolution describes what we see, 
but it also describes what we do {\it not} see - namely, that all branches of the state 
vector (representing all possible measurement outcomes) survive the measurement 
process.  This is nevertheless consistent because it also predicts that an observer 
can be aware of only one such outcome \cite{observer}.  We shall refer to this position 
as unitary quantum theory (UQT).  It includes the Many Worlds Interpretation \cite{Weinberg2,Everett,DeWitt1973}, which asserts that the unobservable branches 
are just as real as the branch we experience, but it is broader.  It includes orthodox 
decoherence theory \cite{Zurek2003,Schloss.19}, whose practitioners represent a 
variety of interpretations \cite{Joos.et.al,Zurek2009}, and other operational 
approaches which assert independence from interpretations \cite{MGM,CerfAdami}, 
while assuming unitarity.

A third position holds that the unobserved branches are removed from the 
theory by a mechanism of yet unknown origin, which takes effect in sufficiently large 
systems, and which is, in principle, subject to quantum analysis.  The mechanism is
represented by adding a nonlinear stochastic term to the Hamiltonian, whose 
effect is to remove all but a single branch \cite{GRW, Pearle,Bassi1,JZchapter8}.  
This approach, in effect, replaces the collapse postulate with an expansion of the 
dynamics postulate beyond its otherwise unitary and deterministic character.  This 
has consequences, which are measurable in principle, but to date undetected.
Predicted effects are difficult to separate from decoherence and other random 
influences.  There are proposals to utilize molecular interferometry and 
optomechanical phenomena, as well as particle diffusion \cite{diffusion}, and 
it is hoped that over the next decade or two, definitive tests will be possible 
\cite{Bassi2}.  We refer to this general position as objective collapse theory (OCT).  
It has fewer adherents than the other two \cite{survey}, but it provides an important 
alternative.

In this paper we will make several points about the measurement process, mostly
interpretation independent, although the blindness of ordinary measurements to
superpositions calls for specific justification within the UQT approach.  For all points it 
will be useful to distinguish two stages of the measurement process.  First comes the 
reversible premeasurement stage, where the object of interest becomes entangled 
(unitarily) with an ancillary system in the apparatus (in our case, two paths).  Second 
is the detection stage, where the ancillary system transfers its entanglement to the 
detector system (in our case, two detectors), which then act irreversibly and record 
a result.   Perhaps surprisingly, two signature collapse properties are established 
in the premeasurement stage, as properties of the object/ancillary system, and survive 
through the detection stage as correlations among the object and the two detectors.  
These are implicit in the postulate stated above:  (i) {\it singleness} (or mutual exclusivity 
of outcomes  $a_n$); and (ii) {\it projection} (the correlations between detector readings 
and the post-measurement spin state).  Note that the singleness of outcomes implies 
randomness.   Randomness is {\it not} a property of the premeasurement state - it only 
shows up at the detection stage as a result of local (``which path'') measurement, which 
breaks the entanglement while preserving the correlations.  There is a {\it third} property 
of the premeasurement state ({\it not} a collapse property!), namely (iii) {\it superposition} 
(the state vector is a superposition of distinct collapse scenarios).  In the ancilla model, 
this property can be detected, but in an ordinary apparatus it cannot be, so that its 
survival in the state vector is open to interpretation. 

The ancilla model expands the premeasurement stage by adding a physical
realization (qubits) to the ancillary system.  The three properties are represented 
by Hermitian operators in the Hilbert space of the object/ancilla system.  The 
operators commute, and all three properties are observable at the detection stage.  
In an ordinary apparatus without the ancillae, we will show that all three properties 
are again present in the premeasurement stage, but that only the two collapse 
properties are observable at the detection stage.  

Interpretations differ on the reason behind this blindness of an ordinary apparatus 
to superpositions.  In two of the approaches outlined above, only one branch of the 
state vector survives the detection stage - in SQT this is axiomatic; in OCT it is by 
construction of the model interaction.  In UQT, on the other hand, the superposition 
extends to the detectors and persists through the detection stage.  We will offer a 
physically intuitive explanation why it is nonetheless undetectable, prompted by 
comparison with the ancilla model.  This will provide a useful perspective on a 
more conventional explanation in decoherence theory (Sec. III and Appendix A).
 

 


In the next section we introduce the Stern-Gerlach measurement model with ancilla qubits
as virtual detectors, and we then show how this atom/ancilla system may be ``observed'' in 
a real experiment.  This observation demonstrates the compatibility of the two collapse 
properties with the superposition property for this system.  In Section III, we compare the 
analogous measurements made with two ordinary detectors.  We show that the collapse 
properties are identical to those of the ancilla model, while the superposition that persists 
in the UQT approach is now undetectable.   We discuss the reason for this blindness and 
compare with the decoherence perspective.
Results are summarized in Sec. IV.

\section{ A Model Measurement with Virtual Detectors}

\begin{figure}[h!]
\includegraphics[scale=0.65]{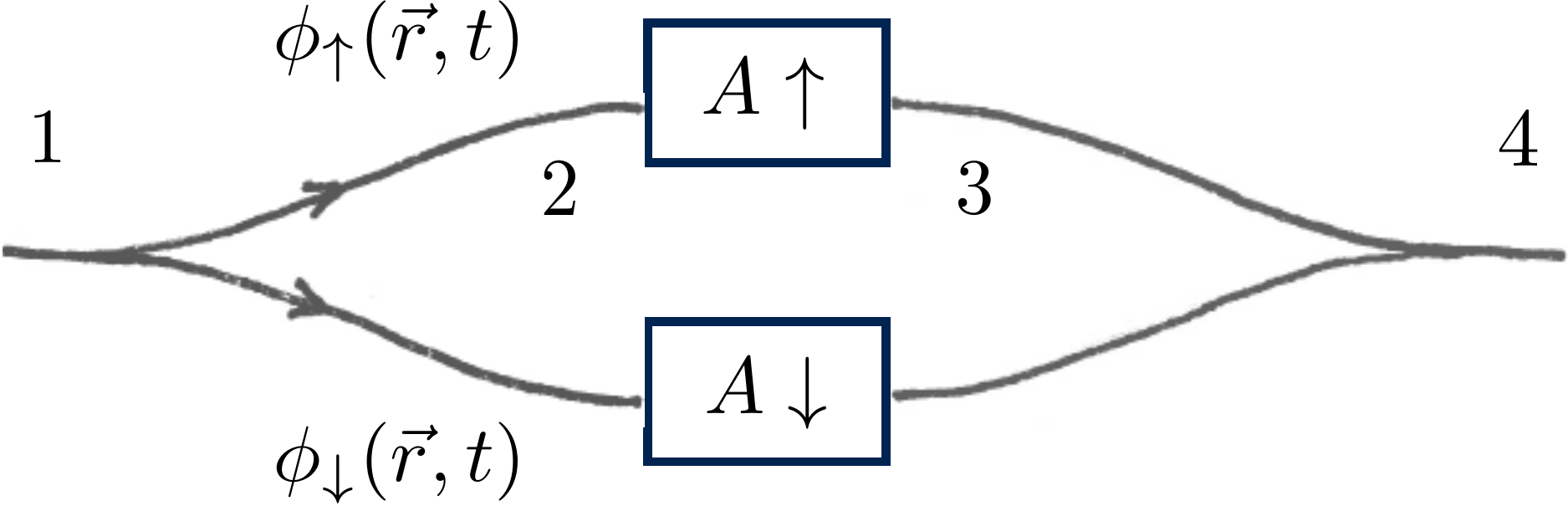}
\caption{\label{fig1} Stern-Gerlach interferometer showing the evolution described by
Eqs. \ref{state1} - \ref{state4}.}
\end{figure}
Consider the compound system consisting of a spin-1/2 atom and two quantum two-state 
systems (ancillae, $A_\uparrow$ and $A_\downarrow$) serving as virtual detectors in the 
Stern-Gerlach interferometer pictured in Fig. 1 \cite{SG.interf.19}.  Each ancilla interacts locally 
with the atom, and it makes a transition from its 0 to its 1 state if and only if the atom passes 
through it.  The interaction is spin-independent, preserving the spin state of the atom on its
path.  We assume that the process is reversible, so that the ancilla by itself does not perform 
a measurement - hence we call it a virtual detector.

Let us trace the evolution of entanglement as the atom passes through the device from 
points 1 - 4.  The atom enters the picture at time $t_1$ with the spatial wavefunction 
$\phi(\vec{r},t_1)$, in an arbitrary pure spin state, $(\alpha,\beta)$,
\be
      \ket{\psi(t_1)} = \phi(\vec{r},t_1) \bigg(\alpha \ket{\uparrow}_s +\beta \ket{\downarrow}_s 
      \bigg) \ket{0}_{A\uparrow} \ket{0}_{A\downarrow},   
\label{state1} 
\ee
with ancillae in their 0 states.  By the time $t_2$, the Stern-Gerlach magnetic field 
gradient has separated the spin components into two ideally nonoverlapping paths, 
\be
      \ket{\psi(t_2)} =  \bigg( \alpha \phi_{\uparrow}(\vec{r},t_2) \ket{\uparrow}_s 
      + \beta \phi_{\downarrow} (\vec{r},t_2) \ket{\downarrow}_s \bigg)   \ket{0}_{A\uparrow} 
       \ket{0}_{A\downarrow}, 
\label{state2} 
\ee
entangling the atom's spatial and spin degrees of freedom.  By $t_3$, the atom has passed 
through an ancilla, $A_{\uparrow}$ {\it or} $A_{\downarrow}$, conditioned on its path, so
that
\be
      \ket{\psi(t_3)} = \alpha \phi_{\uparrow}(\vec{r},t_3) \ket{\uparrow}_s \ket{1}_{A\uparrow}
       \ket{0}_{A\downarrow} + \beta  \phi_{\downarrow}(\vec{r},t_3) \ket{\downarrow}_s
       \ket{0}_{A\uparrow} \ket{1}_{A\downarrow}.    
\label{state3}   
\ee 
Finally, a reversed magnetic field gradient brings the two paths back together at $t_4$.  
Assuming that there is no net phase difference between the paths, the result is
\be
   \ket{\psi(t_4)} =  \phi (\vec{r},t_4) \bigg( \alpha \ket{\uparrow}_s 
      \ket{1}_{A\uparrow} \ket{0}_{A\downarrow} + \beta \ket{\downarrow}_s 
      \ket{0}_{A\uparrow} \ket{1}_{A\downarrow} \bigg).      
\label{state4} 
\ee
The last step allows us to ignore the spatial part and study the remaining entanglement 
between the spin and the two ancillae.  We further simplify by setting 
$\alpha = \beta = 1/\sqrt{2}$, leaving the three-qubit state,
\be
    \ket{\psi(t_4)} \rightarrow \frac{1}{\sqrt{2}} \bigg(\ket{110} + \ket{001} \bigg),
\label{ESG4}
\ee
where the spin states ($\uparrow,\downarrow$) are relabeled as ($1,0$), and the 
ordering of the indices identifies with (spin,$A_ {\uparrow},A_{\downarrow}$).  

This is a Greenberger-Horne-Zeilinger (GHZ) state \cite{GHZ,Mermin.90}, an entangled 
state of three qubits.   It was first realized experimentally in 1999 (\cite{Bouwmeester}) as 
a polarization state of three photons.  Analogous states (and their generalizations to more 
than three particles) have been produced and documented in other systems - for example, 
trapped ions \cite{Monz}, superconducting circuits \cite{Schoelkopf, Song}, and Rydberg 
atoms \cite{Rydberg}.  The original goal was to demonstrate non-locality \cite{nonlocality}; 
more practical goals involve quantum error correction \cite{Shor} and quantum 
communication \cite{QSS}, and in general the manipulation of entanglement.   

Let us discuss the observables that characterize the state, and then the question of how 
to measure them.  The three-qubit system lives in a Hilbert space of dimension eight, and 
$\ket{\psi(t_4)}$ is an eigenstate of three tensor product operators, whose eigenvalues
determine it completely.  The choice is not unique \cite{CSCO}; the most revealing in the 
present context is $ZZI$, $ZIZ$, and $XXX$.   Recalling the definitions of the individual 
Pauli matrices $Z$, $X$, and $Y$, as given in Table I, one may confirm that 
$\ket{\psi(t_4)}$ is indeed the (unique) simultaneous eigenstate of the three tensor 
products with the eigenvalues quoted in Table II.   

Each observable may be characterized by a statement about its physical meaning, with
eigenvalue ($\pm1$) giving the truth value \cite{Zeilinger}.  With GHZ entanglement, all 
such statements concern either two-particle or three-particle correlations, and none 
concerns a property of an individual particle.  The combination of statements appearing 
in Table II is an example.  These statements may appear contradictory, because the 
product of the first two tells us that $IZZ = -1$, whose clear physical meaning is that 
the atom can be found on one and only one path (the ``singleness'' property, or mutual 
exclusivity).  On the other hand, the definiteness of $XXX$ implies that the state vector 
(\ref{ESG4}) is a superposition of two classically inconsistent scenarios.  We will explain 
in detail why the singleness and superposition statements are {\it not} contradictory.  
But first we must describe the measurements, made with real (irreversible) detecting 
devices \cite{Wheeler78}.  We will describe two modes (called local and joint) which 
differ in the acquisition of local information.  

%
\begin{table}
\caption{Multiplication table for Pauli matrices acting on kets ($\ket{1},\ket{0}$)}
\medskip
\begin{tabular}{|c|cc|}
\hline
\ Pauli \  & \ $\ket{1}$ \  & \ $\ket{0}$ \   \\ \hline
$I$ & \  $\ket{1}$ \ & \  $\ket{0}$ \  \\ 
$Z$ & \  $\ket{1}$ \ &  $-\ket{0}$\  \\ 
$X$ & \  $\ket{0}$ \ & \ $\ket{1}$ \  \\ 
$Y$ &  $ \ i\ket{0}$ \ & \ $-i\ket{1}$ \  \\ \hline
\end{tabular}
\smallskip
\end{table}
\begin{table}
\caption{Tensor product operators, eigenvalues and physical interpretations.}
\medskip
\begin{tabular}{|c|c|c|}
\hline
\ (spin,$A\uparrow$,$A\downarrow$) \  & \  eigenvalue \  & \ meaning \   \\ \hline
$ZZI$ & \  $+1$ \ & \ atom takes upper path if and only if spin is up \  \\ 
$ZIZ$ & \  $-1$ \ & \ atom takes lower path if and only if spin is down \  \\ 
$XXX$ & \  $+1$ \ & \ state is invariant under simultaneous flips ($0 \leftrightarrow 1$) \  \\ \hline
\end{tabular}
\smallskip
\end{table}
%



\bigskip
\centerline{\bf A. Local measurements}
\smallskip
 
%
\begin{figure}[h!]
\includegraphics[scale=0.60]{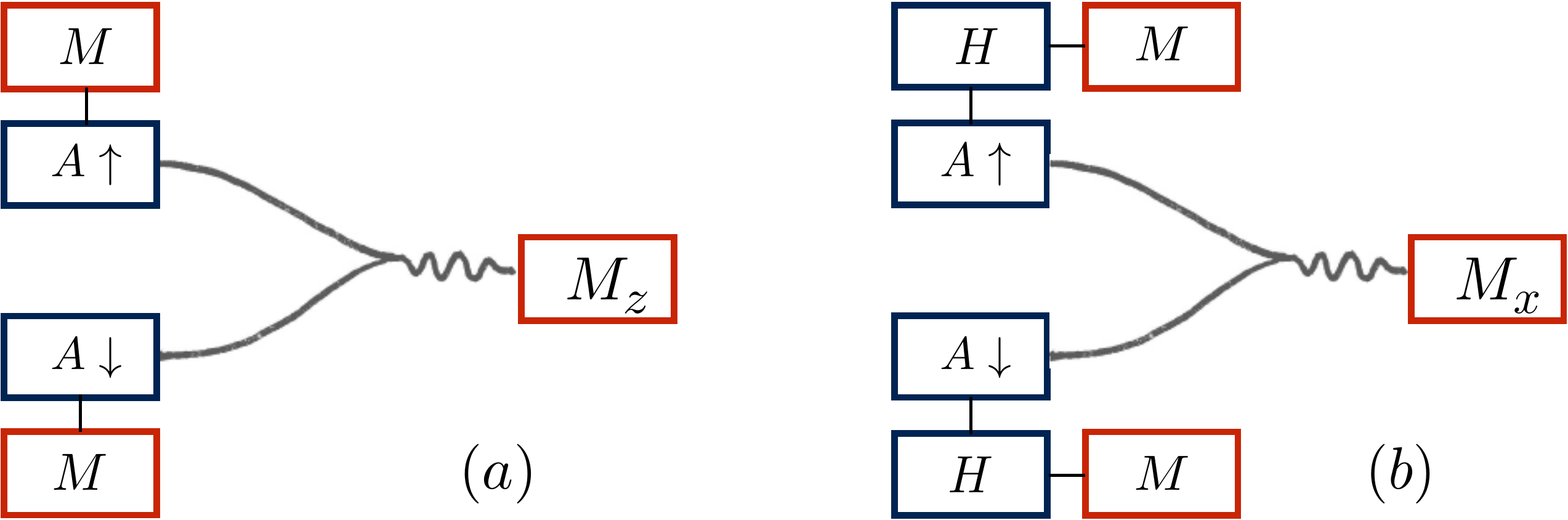}
\caption{\label{fig2} Local measurements of $Z$ factors (a), and $X$ factors (b).  In both
parts, $M$ denotes standard readout devices, and $M_z$ ($M_x$) denote the 
measurement of the atom's spin component $Z_s$ ($X_s$) with a ``downstream'' 
Stern-Gerlach apparatus.  Wavy lines denote the atom's wave packet at $t_4$}
\end{figure}
We begin with the simpler and more typical mode of GHZ experiments 
\cite{Bouwmeester}, in which one measures the local factors and multiplies the results 
together to obtain the eigenvalue of the desired joint observable.  First consider the 
local $Z$ factors, each of which can take the value 1 or $-1$.   We measure the ancilla 
factors, $\ZAup$ and $\ZAdown$, with a standard readout device \cite{readout} for 
each ancilla qubit.  We measure the atomic spin component, $Z_s$, with another 
``downstream'' Stern-Gerlach device, similarly oriented, with a single-atom detector 
placed in the $\uparrow$ path to register the arrival (or not) of the atom and thus 
record the value of $Z_s$.  The results are as follows:   The outcome of each local 
measurement is random because no local observable has $\Psi(t_4) \sim \ket{110} 
+ \ket{001}$ as an eigenstate \cite{randomness}, but the product of {\it any two} 
$Z$ factors is definite, as shown in the top two panels in Table III).  These are collapse 
properties, and they are properties of the state $\Psi(t_4)$ prior to measurement.  They 
characterize the observed collapse phenomenon that we would see in an ordinary 
measurement without the ancillae.   We emphasize that $IZZ = -1$ represents the 
{\it singleness}, or {\it mutual exclusivity} of measurement outcomes, while $ZZI = 1$ 
and $ZIZ = -1$ represent the {\it projection property} - the perfect correlation between 
detector readouts and the (remeasured) value of the atomic spin downstream.   

It is notable that the two-way correlations above persist in the face of local 
measurements ($ZII$, $IZI$, and $IIZ$), whose random outcomes indicate that the 
measurement  changed the state of the system.  The specific ``which path'' information 
thus obtained is {\it not} a property of $\Psi(t_4)$; it enters only at the detection stage.  
Its randomness, a feature shared with ordinary measurements, is a necessary 
consequence of the singleness property.

%
\begin{table}
\caption{Observables characterizing the premeasurement state, and the properties 
they represent.  These properties and their measurements are described in the text.}
\medskip
\begin{tabular}{|c|c|}
\hline
\ observable \  & \ property  \   \\ \hline
$IZZ \rightarrow -1$ \  & \ singleness of outcome \  \\ 
$ZZI \rightarrow 1$ and $ZIZ \rightarrow -1$ \ & \ projection property \ \\
$XXX \rightarrow 1$ and above \ & \ superposition of potential outcomes \  \\  \hline
\end{tabular}
\smallskip
\end{table}

Now consider the measurements of the individual $X$ factors.   Clearly, to measure 
$X_s$, we simply reorient the ``downstream'' Stern-Gerlach system.   To measure 
$\XAup$ and $\XAdown$, noting that the readout devices are keyed to the ancillas' 
$Z$-bases, we apply a Hadamard transformation ($H$) \cite{Hadamard} to 
each ancilla before the readout.  The readout value ($\pm 1$) then indicates in which 
linear combination of $Z$-basis states ($\ket{1} \pm \ket{0}$) each ancilla has been 
``found.''   The results of these $X$-measurements are the following:  The individual
outcomes are again random, but the product of all three is always +1. The definiteness 
of the product indicates a coherent superposition, and its positivity confirms the sign in 
\Eq{ESG4}.   Note that this measurement distinguishes the pure entangled state from 
a mixed state of the same two components.   The mixed state would duplicate the 
$Z$-measurement results but reveal itself through random outcomes for the
product $XXX$.

To better fill out the picture of how the same state, $\Psi(t_4)$, can accommodate both 
a collapse scenario and a superposition, consider the local measurement of the spin 
component $X_s$, which ``finds'' the atom to have taken both paths.   The outcome is 
random:   When it is ($+1$), then the {\it product} $\XAup \XAdown$ must also be 
($+1$).   Knowing the compatible product $\ZAup \ZAdown = -1$ (Table III), it is easy 
to see that the state of the two ancillae is the Bell state, 
\be
    \ket{\Psi}_{AA} \rightarrow {1\over \sqrt{2}} \big( \ket{10} + \ket{01} \big)_{AA}.
\label{ESG5}
\ee
On the other hand, when $X_s$ is ($-1$), then $\XAup \XAdown$ must also be ($-1$), 
and this, combined again with $\ZAup \ZAdown = -1$, indicates another Bell state, 
\be
    \ket{\Psi}_{AA} \rightarrow {1\over \sqrt{2}}  \big( \ket{10} - \ket{01} \big)_{AA}.  
\label{ESG6}
\ee
Each of these Bell states represents perfect anticorrelations between the ancilla's
$Z$-values, which are individually random.  This shows how the ancillae are able to 
respect the singleness property, while (in correlation with the atomic spin) enabling 
the coherent superposition of both paths.  

It may be worth noting that the observation of either superposition state is just
another form of apparent collapse - instead of finding  ``which path,'' one is finding 
``which superposition of paths.'' 

So far, we have appealed to local measurements to verify the joint eigenvalues which
define $\Psi(t_4)$.  But since these measurements change the state, each new set of 
them requires another identically prepared state of the system.  One can circumvent 
this need and achieve a more direct demonstration of compatibility, as follows.  

 \bigskip
 \centerline{\bf B. Joint measurements and compatibility}
 \smallskip
 
%
\begin{figure}[h!]
\includegraphics[scale=0.50]{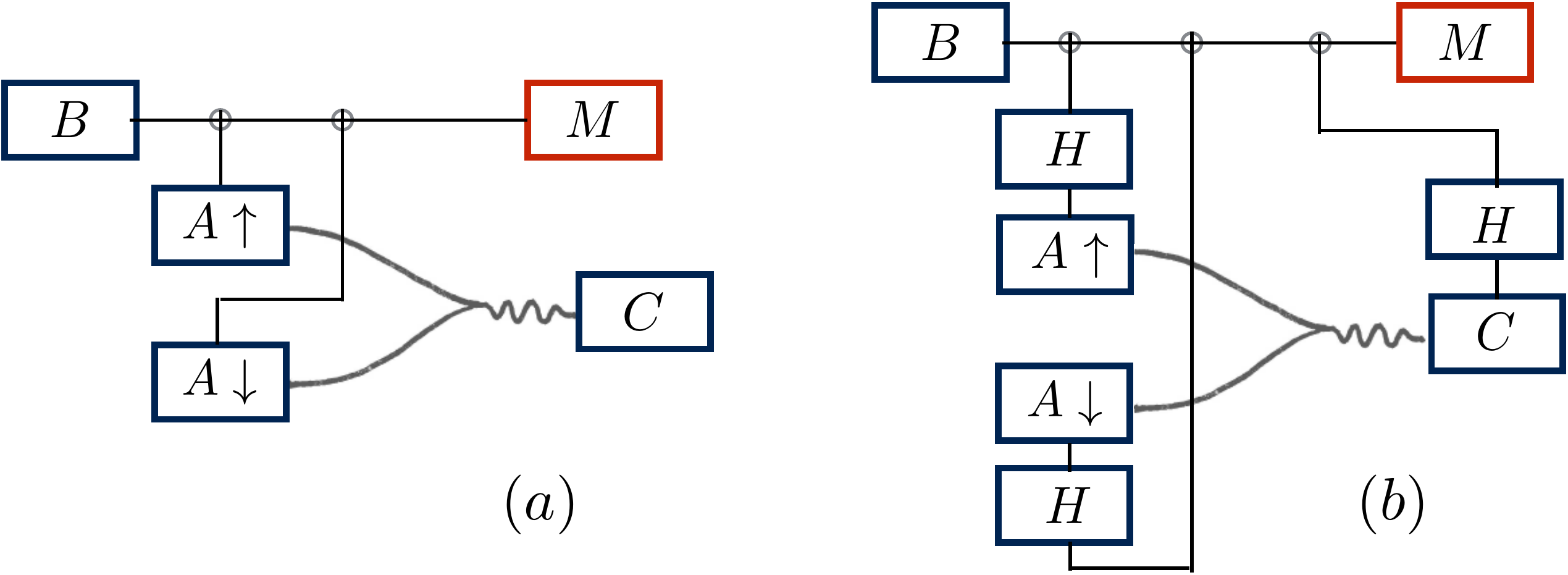}
\caption{\label{fig3} (a) Joint measurement of $IZZ$ using 
ancilla $B$, and (b) of $XXX$ using $B$ and $C$.}
\end{figure}
One can measure compatible joint observables nondestructively by refusing to acquire 
local information.  For the singleness property ($IZZ$), one couples $A_{\uparrow}$ and
$A_{\downarrow}$ to an added ancilla qubit ($B$) through CNOT gates (Fig. 3a), so 
that the state of $B$ changes if and only if $\ZAup$ and $\ZAdown$ are opposite.  
Thus, the readout will show that $IZZ = -1$ without revealing the value of either 
$\ZAup$ or $\ZAdown$.   Similar measurements hold for the projection property 
($ZZI$ or $ZIZ$), by moving one of the CNOT connections to ancilla $C$, located on
the ($\uparrow$) leg of the downstream Stern-Gerlach device.  To measure $XXX$, 
one couples three ancillae ($A_{\uparrow}$, $A_{\downarrow}$, and $C$) to $B$ 
through Hadamard and CNOT gates (Fig. 3b).  The readout will show that $XXX = 1$, 
and this, together with the above, establishes the superposition property.

While these joint measurements provide the (logically) most direct demonstration of  
compatibility, the local measurements have the virtue of demonstrating the randomness 
characteristic of ordinary measurements.   So the ancilla model, subjected to local $Z$
measurements only, is reduced to an ordinary apparatus.   But the $X$ measurements 
give it access to the superposition property, which is {\it not} accessible to ordinary 
measurements.

It should be noted in passing that the ancilla setup in either form can be realized starting
with a photon propagating through a Mach-Zender interferometer and two ancillae, one 
associated with each arm.  This is an extension of the so-called delayed choice 
quantum eraser (DCQE), which employs a single ancilla and is capable of realizing 
Wheeler's proposed delayed choice experiment \cite{Wheeler.DC} (the choice between 
Òwhich pathÓ and a superposition of paths), as closely realized experimentally by 
V. Jacques et. al. \cite{Jacques.07}.  A different realization, and a comprehensive 
discussion of DCQE setups, is provided in \Ref{DCQE}.  With a single ancilla, one has 
a two qubit system and is able to detect the superposition and projection properties 
(XX and ZZ).  Superposition (XX) is manifested by interference as a function of the 
difference in path lengths, and the action of the ancilla X factor corresponds to quantum 
erasure of Òwhich pathÓ information.  But this system cannot detect singleness as the 
third independent observable.  With two ancillae, on the other hand, one has a three 
qubit system which accommodates this observable.  Furthermore, the two-ancillae 
system forms a parallel with an ordinary apparatus which employs two detectors.  We 
show below that the singleness property must hold here as a correlation between 
the detectors.


\section{Apparati with Real Detectors}

Reconsider the Stern-Gerlach setup of Fig. 1 with ancillae replaced by real 
detectors, $\Dup$ and $\Ddown$.  These act like the ancillae together with their 
readout devices:  Like the ancillae, they transmit the atom without changing its spin 
state (this is in keeping with the von Neumann measurement formalism \cite{VN55,Ideal}
and the stated postulate).   And, like the readout devices, they record (irreversibly) the 
passage (or not) of the atom.

In this section, first, we show that the collapse properties of ordinary apparati are 
equivalent to those of the ancilla model, in the sense that the singleness and projection 
properties are represented by observables and established at the premeasurement 
stage, while the randomness of ``which path'' information enters only at the detection 
stage.  Then second, we discuss the superposition property, which survives the detection 
stage in UQT but is nonetheless undetectable.  We offer a physical explanation why this 
is the case.

\medskip
\centerline{{\bf A. Equivalence of Collapse Properties}}
\smallskip

Recall that the ancilla-based premeasurement state, $\Psi(t_4)$, exhibits nonrandom 
collapse properties represented by joint observables:  (i) the singleness property, by 
$\ZAup \ZAdown = -1$, and (ii) the projection property, by $Z_s \ZAup  = +1$ and 
$Z_s \ZAdown  = -1$.  To show that these properties are similarly (pre)established 
in ordinary measurements, note that the ancilla operators, $\ZAup$ and $\ZAdown$, 
act as stand-ins for path occupation variables.  And we can define such variables 
for the ordinary setup as $\ZPup = 2\CP_{\uparrow} - 1$ and 
$\ZPdown = 2\CP_{\downarrow} - 1$, where $\CP_k$ is a projector onto that   
subvolume of the path $k = (\uparrow,\downarrow)$ occupied by the spatial wave 
packet $\phi_k(\vec{r},t_2)$ at time $t_2$.  So $\ZPk \rightarrow \pm 1$ tells us whether 
or not the atom will enter the detector $\Dk$.  The path occupation {\it states} analogous 
to the ancilla states may be written as 
$\ket{1}_{P\uparrow}\ket{0}_{P\downarrow}$ and 
$\ket{0}_{P\uparrow}\ket{1}_{P\downarrow}$.  These convey all of the necessary 
information about the spatial wave packets $\phi_{\uparrow}(t_2)$ and 
$\phi_{\downarrow}(t_2)$, so that $\ket{\psi(t_2)}$ may be written as
\be
  \ket{\psi(t_2)} =  \frac{1}{ \sqrt{2}} \bigg(  \ket{1}_s 
      \ket{1}_{P\uparrow} \ket{0}_{P\downarrow} +  \ket{0}_s 
      \ket{0}_{P\uparrow} \ket{1}_{P\downarrow} \bigg),     
\label{state8} 
\ee
which is analogous to \Eq{state4} of the ancilla system prior to readout.  This is clearly 
an eigenstate of $\ZPup \ZPdown$, $Z_s \ZPup$, and $Z_s \ZPdown$, with eigenvalues 
-1, 1, and -1, respectively, thus establishing the singleness and projection properties in 
the premeasurement  state.  These properties arise here from the entanglement of just 
the spin and the path, with  path occupation alone playing the ancillary role. 

By the time $t_3$, the path occupation variables have mediated correlations between 
the atom's spin and the detectors - analogous to the readout state in the ancilla case. 
The main difference is that here, a detector readout follows closely the passage of 
the atom - one cannot delay this readout as was done with the ancilla system.  But 
the ancilla system can duplicate this situation by moving the ancilla readouts to an
earlier time, say $t_3$, ahead of the atomic spin remeasurement (at some $t > t_4$).  
The time ordering has no effect on the final result \cite{ordering}.  

As an aside on  the projection property - although this property is axiomatic in SQT, it is
nonetheless conditioned on the above assumption that  a detector transmits the atom 
without changing its spin state \cite{Ideal}.  Clearly this assumption fails for detectors 
which work by absorbing the atom.   In this case the projection property still holds, but 
it takes the form of a correlation between the reading of a detector and the angular 
momentum imparted to it.  Only one detector can receive the impulse (the  singleness 
property), and that impulse ($\pm \hbar$) is delivered to $\Dup$ or $\Ddown$, 
respectively, depending on which detector reads 1.   

\medskip
\centerline{{\bf B. Blindness to Superpositions}}
\smallskip

While the collapse properties of the two systems are equivalent, the superposition 
property is not, being detectable in the ancilla system but not in the ordinary system.  
All approaches outlined at the beginning of the paper must agree on the facts:  the 
observation of a single definite outcome (with its associated probability), and the 
invisibility of superpositions to ordinary apparati.  But the approaches differ 
fundamentally on the physics behind these facts.  In the SQT and OCT approaches, 
because the detectors are macroscopic, one of the terms existing in the 
premeasurement state is removed from the theory at detection.  In SQT the removal 
is axiomatic; in OCT it is dynamical, a result of the nonlinear stochastic model.  The 
evolution is nonunitary in both cases.  In the UQT approach, on the other hand, the 
evolution including the detectors is unitary, and therefore, the state of the 
atom/detectors system must reflect the superposition in \Eq{state8}.  This approach 
is viable, as we have said, only if the superposition of distinct detector states is 
undetectable.  Decoherence theory shows that it is (see Appendix A), but the ancilla 
analogy suggests a more fundamental explanation involving irreversibility
\cite{Raimond.97}.   

In UQT, the state of the atom/detectors system, at time $t_4$, has the same GHZ form 
as \Eq{state8} (or, for that matter, \Eq{ESG4} of the ancilla system).  The crucial 
difference is that the detector states involve microscopic internal degrees of freedom 
\cite{micro-degrees}, whose states are labeled by $\mu$ and $\mu'$ in addition to 
the necessary readout variables, 0 and 1 respectively.  So \Eq{state8} becomes
\be
    \ket{\psi(t_4)} =  {1\over \sqrt{2}} \bigg( \ket{1}_s 
      \ket{1,\mu'}_{D\uparrow} \ket{0,\mu}_{D\downarrow} + \ket{0}_s 
      \ket{0,\mu}_{D\uparrow} \ket{1,\mu'}_{D\downarrow} \bigg),    
\label{state9} 
\ee 
where we have dropped the common spatial factor at $t_4$ as done in \Eq{ESG4}.
One can imagine (0) to be a metastable configuration of a detector, which would make 
a transition to a final stable configuration (1) if triggered by the passage of the atom.   
Equation \ref{state9} represents just a single element of an ensemble in which each 
initial state $\mu$ of the (0) configuration evolves unitarily into the state $\mu'$ of the 
(1) configuration \cite{caveat}.  The corresponding density matrix is written in 
Appendix A.  In the text, for clarity, we shall continue to refer to state vectors.

Again we ask - how can one observe a superposition involving both paths?   One must 
measure, among other things, the spin component $X_s$.   Since this measurement by 
itself produces random outcomes, one must measure a correlation of which $\Psi(t_4)$ 
is an eigenstate, of which the simplest is $XXX$.   There are three other options, such 
as $XYY$, but these offer nothing further.  So, supposing that the $X_s$ measurement 
produces the outcome $+1$, we must then show that the product of detector $X$ 
values is also $+1$.  Given that the product of detector $Z$ values is $-1$, this would 
demonstrate that the detectors are in the Bell state analogous to \Eq{ESG5}:
\be
    \ket{\Psi}_{DD} \rightarrow {1\over \sqrt{2}} \bigg( \ket{1,\mu'}_{\Dup} 
    \ket{0,\mu}_{\Ddown} + \ket{0,\mu}_{\Dup} \ket{1,\mu'}_{\Ddown} \bigg).
\label{ESG7}
\ee
A detector operator $X_k$ connects its 0 and 1 states, that is, 
$\ket{1,\mu'}_{Dk} = X_{Dk} \ket{0,\mu}_{Dk}$ and 
$\ket{0,\mu}_{Dk} = X_{Dk} \ket{1,\mu'}_{Dk}$.   Since the first of these represents the 
natural evolution of the detector, $\ket{1,\mu'} = U(t_3,t_2) \ket{0,\mu}$, the 
$X$ operators must be
\be
    X_{Dk} = P_k(1) U_k (t_3,t_2) P_k(0) + P_k(0) U_k^{-1}(t_3,t_2) P_k(1),
\label{Xoperator}
\ee where $P_k(i)$ are projection operators onto the $i = 0$ or 1 configurations of
detector $D_k$.  Now $U_k (t_3,t_2)$ represents the time evolution of a complex 
many-body system, and while this is reversible {\it in principle}, it is not reversible 
{\it thermodynamically} \cite{Batal.15}. That is, we do not have control over the 
microscopic degrees of freedom required to implement $U_k^{-1}(t_3,t_2)$.
So the detector 
operators $X_k$ are not accessible to us, and without them we cannot detect a 
superposition of states in the  0 and 1 configurations of $D_k$ - and we cannot 
access $XXX$, which would demonstrate a \Sch~cat-like superposition of the two 
collapse scenarios of the spin/detectors system.  In short, we only have access to 
the detector variables $Z_k$ which  the detectors record, and in these there can be 
no evidence for the existence of the superposition which (in UQT) continues to exist 
in the state vector.  A different but related argument on thermodynamic irreversibility 
in measurement was given by Peres \cite{Peres80}. 

The demonstration of blindness changes very little with detectors which absorb the 
atom rather than transmitting it, as is essentially the case in the original Stern-Gerlach 
experiment \cite{Stern Gerlach}.  This case is discussed in Appendix B.

The above arguments suggest that our inability to detect superpositions of detector 
output states is a manifestation of the second law of thermodynamics.   One can 
imagine a quantum Maxwell Demon who possesses the microscopic control that we 
lack, and is capable of detecting superpositions which are invisible to us.  Thus, the 
quantum measurement process, by its construction, employs the thermodynamic 
arrow of time.  There is no need to invoke a different ``measurement'' arrow.   

\medskip
\centerline{{\bf C. A Decoherence Perspective}}
\smallskip
A decoherence argument for blindness involves the concept of the pointer;
Brasil \cite{Brasil15} defines ``pointer states (as) eigenstates of the observable 
of the measuring apparatus that represents the possible positions of the display 
pointer of the equipment.''   The concept was introduced in the present context 
by Zeh \cite{Zeh.70} and developed by Zurek \cite{Zurek.81,Zurek.82}, who showed 
that interactions with the environment select the pointer's preferred basis (the
states we observe).  In our system these states are associated with the {\it pair} 
of detectors; they are denoted by $\ket{1}_{D\uparrow} \ket{0}_{D\downarrow}$ and 
$\ket{0}_{D\uparrow} \ket{1}_{D\downarrow}$ (or in shorthand, $\ket{10}_{DD}$ 
and $\ket{01}_{DD}$), by simply dropping the environmental variables $\mu$ 
and $\mu'$.

An argument specific to our system is written out in Appendix A. In brief, one 
traces over the environmental degrees of freedom ($\mu$ and $\mu'$) within 
each detector, and derives the reduced density matrix of the spin/pointer system, 
which is diagonal in the basis $\ket{110}_{sDD}$ and $\ket{001}_{sDD}$, where 
the second and third indices refer to the pointer.  In the case of detectors which 
absorb the atom, this basis reduces to just that of the pointer, $\ket{10}_{DD}$ 
and $\ket{01}_{DD}$.  In either case, blindness to superpositions is 
represented by the diagonality of the ($2 \times 2$) density matrix. 

There are interesting consequences when the pointer is realized within the system 
of two detectors.  Environmental interactions determine the preferred basis of an
individual detector, $\ket{1}_{Dk}$ and $\ket{0}_{Dk}$ 
(corresponding to {\it atom/no atom}, or $Z_{Dk} = \pm 1$).  But they do {\it not}
determine the preferred basis of the full pointer associated with the {\it pair} of 
detectors, because they do not exclude the possibility of $\ket{11}_{DD}$ and 
$\ket{00}_{DD}$ - exclusion resulting from the singleness property.  In fact, the 
environment is not responsible for the singleness or the projection property, or 
implicitly, for the choice of which spin component is measured - all of which are 
established in premeasurement.  So, while the environment is crucial for enforcing 
the blindness to superpositions, its role (whether internal or expanded to include 
the external) is limited to the proper functioning of the individual detectors.

This separation between the premeasurement stage (governed by reversible 
unitary evolution), and the detection stage (where the environment enters 
bringing practical irreversibility), is the defining characteristic of controlled von 
Neumann-type measurements.  The roles of these stages are in some sense 
complementary:  While blindness to superpositions may be seen as an emergent 
classical property induced by interactions with the environment, the singleness 
and projection properties are quantum entanglement properties, represented by 
observables and manifested in correlations between noninteracting macroscopic 
objects.

\section{Conclusions}

We studied an ancilla-aided Stern-Gerlach experiment allowing delayed-choice 
measurements on the three-qubit system of atomic spin and two ancillae acting 
as virtual detectors.  We first considered local measurements, and showed that 
one choice ($Z_i$) reproduces the collapse properties of ordinary (unaided) 
measurements, while another ($X_i$) demonstrates a superposition of the two 
collapse-like scenarios, involving both paths.  Both choices require repeated 
measurements on identically-prepared states of the system, since local 
measurements destroy the state.

So secondly, we showed that nondestructive measurements can be made by 
avoiding the acquisition of local information.  Thus relinquishing only the specific 
``which path'' information, one can still measure a complete set of commuting
joint observables - these represent the superposition property and the two 
(nonrandom) collapse properties, namely the singleness and projection properties. 

In Sec. III we applied the above ideas to an ordinary apparatus.  First, we showed 
that the collapse properties occur with the same status as in the ancilla model:  
The singleness and projection properties are represented by operators and 
established at the premeasurement stage, while the randomness of local
measurement outcomes enters only at the detection stage, as a consequence of
the singleness property.  The various approaches mentioned in the introduction 
- the standard, unitary, and objective collapse approaches - agree on the 
observed collapse phenomenon itself, but they differ on its unobservable 
underpinnings - the existence/nonexistence of unobserved branches in the 
state vector - and the nature of the observed randomness of outcomes (is it 
objective or subjective?).  It is possible, but far from certain, that future
experiments alone will resolve these differences.

The viability of the UQT approach rests upon the invisibility of the alternate
(unobserved) branches in the state vector.   The ancilla system points to a 
simple explanation:  Detectors are irreversible, and this makes the required 
complementary local observables ($X_{D\uparrow}$ and $X_{D\downarrow}$) 
inaccessible.  The absence of a known fundamental mechanism of irreversibility
acting in typical measurements suggests that the irreversibility is thermodynamic, 
so that (at least within UQT), the  observed collapse phenomenon is a 
manifestation of the second law.  

\medskip
\centerline{{\bf Appendix A:  Density Matrix of Spin-Detectors System}}
\smallskip

Here we write out the density matrix of the spin/detectors system, and we show 
how the trace over the unobserved states of the detectors' internal degrees of 
freedom yields the appropriate reduced density matrix, which expresses the 
blindness of the apparatus to superpositions of output states. 

The initial mixed state of the spin/detectors system, assuming probabilities 
$p_{\muup}$ and $p_{\mudown}$ for the microstates, $\ket{0,\muup}_{\Dup}$ 
and $\ket{0,\mudown}_{\Ddown}$ of the two detectors, is
\be
   \rho(t_1) =  \sum_{\muup,\mudown} p_{\muup} p_{\mudown} 
   \bigg( \alpha \ket{1}_s +\beta \ket{0}_s \bigg) \ket{0,\muup}_{\Dup} 
   \ket{0,\mudown}_{\Ddown} \bigg( \alpha^* \bra{1}_s +\beta^* \bra{0}_s \bigg)
   \bra{0,\muup}_{\Dup} \bra{0,\mudown}_{\Ddown}. 
\label{AppB1}
\ee
After the atom passes through the detectors and the paths are recombined at 
$t_4$, this becomes
\bea
   \rho(t_4) =  \sum_{\muup,\mudown}  &  p_{\muup} p_{\mudown} 
   \bigg( \alpha \ket{1}_s \ket{1,\muup'}_{\Dup} \ket{0,\mudown}_{\Ddown}  + 
   \beta \ket{0}_s \ket{0,\muup}_{\Dup} \ket{1,\mudown'}_{\Ddown} \bigg) \cdot \nonumber  \\
    & \cdot \bigg( \alpha^* \bra{1}_s \bra{1,\muup'}_{\Dup} \bra{0,\mudown}_{\Ddown} + 
   \beta^* \bra{0}_s \bra{0,\muup}_{\Dup} \bra{1,\mudown'}_{\Ddown} \bigg).
\label{AppB2}
\eea
Since we only read the detectors' outputs and do not monitor the microscopic degrees 
of freedom (considered as the ``environment'' $E$), we trace over the latter to define 
the reduced density matrix describing the state of the spin and the detector displays, 
which is called the spin/pointer system \cite{Brasil15}).  To be more precise, 
each detector consists of its own pointer (with readout states 0 and 1), 
and its own environment (with associated states $\mu$ and $\mu'$, 
respectively).  The trace consists of independent traces over the environments within
each detector, \ie, $\rho^r(t_4) \equiv Tr_{\Eup,\Edown} \rho(t_4)$.  To evaluate each 
of these, it is convenient to sum over $\mu$ (\ie, $\sum_\mu \bra{\mu} ... \ket{\mu}$ 
in those terms where the pointer state 0 appears, and over $\mu'$ where it does not.  
The latter choice is legitimate because the two sets are related unitarily.  It is 
straightforward then to show that
\bea
  & \rho^r(t_4)  =   |\alpha|^2 \ket{1}_s \ket{1}_{\Dup} \ket{0}_{\Ddown} 
   \bra{1}_s \bra{1}_{\Dup} \bra{0}_{\Ddown}   \nonumber  \\
   &  + \alpha \beta^* \ket{1}_s \ket{1}_{\Dup} \ket{0}_{\Ddown} \bra{0}_s  
   \bra{0}_{\Dup} \bra{1}_{\Ddown} \sum_{\muup,\mudown} p_{\muup} p_{\mudown} 
   \braket{\muup}{\muup'}_{\Dup} \braket{\mudown'}{\mudown}_{\Ddown} \nonumber \\
   &  + \alpha^* \beta \ket{0}_s \ket{0}_{\Dup} \ket{1}_{\Ddown} \bra{1}_s  
   \bra{1}_{\Dup} \bra{0}_{\Ddown} \sum_{\muup,\mudown} p_{\muup} p_{\mudown} 
   \braket{\muup'}{\muup}_{\Dup} \braket{\mudown}{\mudown'}_{\Ddown} \nonumber \\
   &  + |\beta|^2 \ket{0}_s \ket{0}_{\Dup} \ket{1}_{\Ddown} 
   \bra{0}_s \bra{0}_{\Dup} \bra{1}_{\Ddown} 
\label{reduction}
\eea
The environmental sums in the second and third terms essentially vanish (they are
undetectably small) because the inner product factors, none greater than unity in 
magnitude, have random phases, in contrast with analogous factors
$\big(\braket{\mu}{\mu}\braket{\mu'}{\mu'} = 1\big)$ which appeared in the first and 
fourth terms and summed to unity.  Thus $\rho^r$ is diagonal in the spin-pointer 
basis, which consists of $\ket{1}_s \ket{1}_{\Dup} \ket{0}_{\Ddown}$ and 
$\ket{0}_s \ket{0}_{\Dup} \ket{1}_{\Ddown} $.   The surviving singleness and 
projection correlations result from the entanglement generated between 
\Eqs{AppB1}{AppB2} by the passage of the atom.   In fact it should be noted that, 
except for the remaining summations in the off-diagonal terms, \Eq{reduction} is 
equivalent in form to the density matrix of the spin-ancilla system (see \Eq{state4} 
in Sec. II).  Thus, the environmental factors in \ref{reduction} neatly summarize 
how the superposition of outcomes becomes undetectable with real detectors.
  
\medskip
\centerline{{\bf Appendix B: Detectors that Absorb}}
\smallskip

In the original Stern-Gerlach experiment \cite{Stern Gerlach}, silver atoms were 
directed at a glass plate and formed two separated deposits, with segments of the 
glass acting as the two detectors.   Imagining ideally a pair of absorbing single-atom 
detectors, their state at time $t_3$ could still be written as in \Eq{ESG7}, but the 1 
states now represent the absorbed atom as well as excitations created by the 
absorption event.   Natural evolution produces these states from the 0 states of the 
detectors multiplied by the corresponding path occupation states $\ket{1}_{Pk}$ of 
the atom:  $\ket{1,\mu'}_{Dk} = U(t_3,t_2) \ket{0,\mu}_{Dk} \ket{1}_{Pk}$. 
So the $X_{Dk}$ operator analogous to \Eq{Xoperator} is  
\be
    X_{Dk} = P_k(1) U_k (t_3,t_2) \ket{1}_{Pk} P_k(0) \bra{1}_{Pk}  
    + \ket{1}_{Pk} P_k(0) \bra{1}_{Pk}  U_k^{-1}(t_3,t_2) P_k(1),
\label{Xoperator2}
\ee
and the subsequent blindness argument is unchanged.   

The decoherence approach of Appendix A is similarly adapted:  Since 
the $\ket{\mu'}$ states include the absorbed atom, the inner product factors  
in \ref{reduction} are replaced by $\bra{1}_{Pk} \braket{\mu}{\mu'}$ or its 
complex conjugate.  The set \{$\mu'$\} is not complete because it refers to more
particles than $\{\mu\}$, but it includes all states generated unitarily from \{$\mu$\} 
and the incident atom.  

%

\begin{thebibliography}{99}
\bibitem{preMGM} There are more general statements of this postulate, extending  
      beyond orthogonal measurements and ideal measurement conditions.  See the 
      comprehensive discussion in \Ref{MGM}.
\bibitem{Ideal} An ideal measurement, as employed by von Neumann \cite{VN55} and 
       defined in \Ref{Pokorny}, is projective.
\bibitem{relevant}  In principle, the environment could be the rest of the universe, but Zurek
      discusses the more realistic notion of that part most responsible for decoherence, defined 
      within the appropriate closed system (see \Ref{Zurek.81}, p. 1520).  In this work we will
      regard the relevant environment to be just the internal degrees of freedom of the 
      individual detectors.
\bibitem{CSCO} Mermin makes a different choice in his analysis of non-locality and 
      contextuality in a similar GHZ state (\Ref{Mermin.90}, p. 3375, line 1).  The three 
      observables must generate the stabilizer group of the state, and any choice which 
      includes a three-particle correlation will do this.
\bibitem{randomness} If a state is completely determined by correlation properties, then
      it cannot possess local properties.  GHZ states and Bell states are classic examples.
\bibitem{ordering} It is interesting to note that a spacelike separation of local measurement 
       events is possible, in principle.  This rules out causal connections between them.
\bibitem{micro-degrees} The internal degrees of freedom of the individual detectors are 
       sufficient to produce decoherence (see Appendix A), and hence they comprise the
       relevant environment \cite{relevant}.  External degrees of freedom could contribute, 
       but not decisively.  And they certainly do not mediate interactions between 
       the detectors.
\bibitem{caveat}  To define $\mu'$ more precisely, it is the microscopic state of the (1)
       configuration at the time when the outcome (1) is recorded.  This is the time when 
       the measurement is completed  \cite{Wheeler78}.
\end{thebibliography}
\end{document}